\documentclass{article}
\usepackage{cite}
\usepackage{amsmath}
\usepackage{amsfonts}
\usepackage{amssymb}
\usepackage{amsthm}

\def\su{\mathfrak{su}}
\def\sh{\mathfrak{sh}}
\def\sl{\mathfrak{sl}}

\newcommand{\be}{\begin{equation}}
\newcommand{\ee}{\end{equation}}

\title{Dynamical symmetry of a semiconfined harmonic oscillator model with a position-dependent effective mass}

\author{E.I. Jafarov
\thanks{\emph{Corresponding author:} E-mail: ejafarov@physics.science.az}%
\and S.M. Nagiyev
\thanks{E-mail: sh.nagiyev@physics.science.az}
\\ Institute of Physics, State Agency for Science and Higher Education \\ Javid ave. 131, AZ1143, Baku, Azerbaijan
}                     
\begin{document}

\maketitle
\begin{abstract}
Dynamical symmetry algebra for a semiconfined harmonic oscillator model with a position-dependent effective mass is constructed. Selecting the starting point as a well-known factorization method of the Hamiltonian under consideration, we have found three basis elements of this algebra. The algebra defined through those basis elements is a $\su\left(1,1 \right)$ Heisenberg-Lie algebra. Different special cases and the limit relations from the basis elements to the Heisenberg-Weyl algebra of the non-relativistic quantum harmonic oscillator are discussed, too.
\vspace{2ex}

\noindent
{\bf 2020 Mathematics Subject Classification.} Primary: 12D05, 12H05, 33C45; Secondary: 34L40, 47A68.
\vspace{2ex}

\noindent
{\bf Keywords:} Position-dependent effective mass; Quantum harmonic oscillator; Semiconfinement effect; Heisenberg-Lie algebra; Exact expression.
\end{abstract}

\section{Introduction}

The harmonic oscillator is the most attractive problem of quantum mechanics without any doubt~\cite{landau1991,moshinsky1996,bloch1997}. This attractivity covers not only theoretical and experimental studies but also mathematical foundations. The non-relativistic quantum harmonic oscillator model constructed within the canonical approach exhibits main characteristics in terms of its wavefunctions of the stationary states expressed through the Hermite polynomials, discrete energy spectrum consisting of the infinite number equidistant energy levels with a ground-state level being equal to $\frac 12 \hbar \omega$ as well as the dynamical symmetry algebra, where the position and momentum operators together with the Hamiltonian $\hat H$ form the Heisenberg-Weyl dynamical symmetry algebra. The Hamiltonian further can be factorized in terms of the harmonic oscillator creation and annihilation operators $\hat a^\pm$. These two operators are sufficient for the extraction of all necessary useful information from the non-relativistic quantum harmonic oscillator without a direct solution of the Schr\"odinger equation corresponding to it.

There are a lot of exactly-solvable generalizations of the above-described oscillator model within the various quantum-mechanical approaches. From the mathematical viewpoint, the listed above characteristics for some of them can overlap, but not in the case of their wavefunctions of the stationary states. These generalized wavefunctions have analytical expressions in terms of various orthogonal polynomials belonging to the Askey scheme and its $q$-deformed version. As we are aware, there exists only one more quantum harmonic oscillator model, which dynamical symmetry algebra completely overlaps with the dynamical symmetry algebra of the non-relativistic quantum harmonic oscillator model constructed within the canonical approach -- it has the Heisenberg-Weyl dynamical symmetry algebra. This is a discrete quantum harmonic oscillator model with the wavefunctions of the stationary states expressed through the Charlier polynomials~\cite{atakishiyev1989,atakishiyev1998}. Motion of the non-relativistic quantum harmonic oscillator under the suddenly exposed constant external field is an exception here. The rest of the known exactly-solvable quantum harmonic oscillator models exhibit dynamical symmetries, which belong to Lie (super)algebra or $q$-deformation of the Heisenberg-Weyl algebra~\cite{green1953,ryan1963,atakishiev1980,ganchev1980,ohnuki1982,macfarlane1989,kagramanov1990,atakishiev1990,floreanini1991,vanderjeugt1992,atakishiyev2001,jafarov2008,jafarov2011a,jafarov2011b,jafarov2012a,jafarov2012b,jafarov2013,belmonte2020}. 

Recently, \cite{jafarov2021} discussed the problem of the exactly-solvable quantum harmonic oscillator within the non-relativistic canonical approach, which exhibits a semiconfinement effect. Its wavefunctions of the stationary states vanish at both values of the position $x=-a$ and $x \to +\infty$. Such behavior is achieved thanks to the introduction of the position-dependent effective mass instead constant one~\cite{mathews1975,carinena2004,schmidt2007,amir2014,quesne2015,karthiga2017,carinena2017,carinena2019,jafarov2020a,jafarov2020b,jafarov2020c,jafarov2021a,jafarov2021b}. Wavefunctions of the stationary states of the model are expressed via the generalized Laguerre polynomials, but, its energy spectrum completely overlaps with the energy spectrum of the non-relativistic canonical quantum harmonic oscillator described above. Later, this quantum oscillator model was also generalized to the case of the applied external homogeneous field~\cite{jafarov2022} as well as its phase space was constructed and the exact expression of the Husimi function of the joint quasiprobability of the momentum and position was computed in~\cite{jafarov2022b}. Despite the fact that the semiconfinement oscillator model was developed in-depth within these published papers, one important building block of the model is still absent -- its dynamical symmetry algebra is not discussed at all. In the present paper, we complete the absent block and show that the semiconfined oscillator model also has elegantly described dynamical symmetry algebra. 

Our paper is structured as follows: Section 2 consists of the known information about the standard nonrelativistic canonical quantum harmonic oscillator and its Heisenberg-Weyl dynamical symmetry algebra. Section 3 is devoted to the construction of the dynamical symmetry algebra of the oscillator model with a position-dependent effective mass exhibiting semiconfinement effect. The final section is devoted to detailed discussions. It also includes limit relations between the dynamical symmetry algebra of the oscillator model with a position-dependent effective mass exhibiting semiconfinement effect and the Heisenberg-Weyl dynamical symmetry algebra of the standard nonrelativistic canonical quantum harmonic oscillator.

\section{The Heisenberg-Weyl algebra of the non-relativistic quantum harmonic oscillator within the canonical approach}

The content of the current section is more informative because all expressions presented here exist in most quantum mechanics textbooks and review papers. We include here straightforward computations regarding the Heisenberg-Weyl algebra of the non-relativistic quantum harmonic oscillator within the canonical approach, because, a generalized approach for the construction of the dynamical symmetry algebra of the oscillator model with a position-dependent effective mass exhibiting semiconfinement effect will follow by some a way the method provided below. 

It is well-known that the wavefunctions of the stationary states and the energy spectrum of the non-relativistic quantum system within the canonical approach can be obtained exactly from the one-dimensional time-independent Schr\"odinger equation

\be
\label{sheq-gen2}
\left[ { - \frac{{\hbar ^2 }}{{2m_0 }}\frac{{d^2 }}{{dx^2 }} + V\left( x \right) - E_n } \right]\psi _n \left( x \right) = 0, \quad n = 0,1,2, \ldots,
\ee
where $m_0$ is a constant mass of the quantum system under consideration.

\cite{infeld1951} shows that the Hamiltonian of eq.(\ref{sheq-gen2}) can be easily factorized as follows

\be
\label{h-f}
\hat H = \hbar \omega \hat a^ +  \hat a^ -   + E_0 ,
\ee
in terms of two first-order differential operators $\hat a^ +$ and $\hat a^ -$ of the following mathematical expressions:

\begin{align}
\label{a+}
\hat a^ +   =&  - \frac{1}{{\sqrt 2 \lambda _0 }}\frac{d}{{dx}} + W\left( x \right), \\
\label{a-}
\hat a^ -   =& \frac{1}{{\sqrt 2 \lambda _0 }}\frac{d}{{dx}} + W\left( x \right).
\end{align}
where $\lambda _0  = \sqrt {\frac{{m_0 \omega }}{\hbar }}$.

Usually, $W\left( x \right)$ introduced above is called a superpotential, and the potential energy $V\left( x \right)$ can be easily expressed through it as follows:

\[
V\left( x \right) = E_0  + \hbar \omega W^2 \left( x \right) - \frac{{\hbar \omega }}{{\sqrt 2 \lambda _0 }}W'\left( x \right).
\]

One can easily check that the following commutation relations between Hamiltonian $\hat H$ defined through (\ref{h-f}) and operators $a^ \pm$ defined through (\ref{a+}) and (\ref{a-}) hold:

\begin{align}
\label{ha+}
 \left[ {\hat H,\hat a^ +  } \right] =& \sqrt 2 \frac{{\hbar \omega }}{{\lambda _0 }}W'\left( x \right)\hat a^ +   - \frac{{\hbar \omega }}{{\lambda _0 ^2 }}W''\left( x \right), \\ 
\label{ha-}
 \left[ {\hat H,\hat a^ -  } \right] = & - \sqrt 2 \frac{{\hbar \omega }}{{\lambda _0 }}W'\left( x \right)\hat a^ -  , \\ 
\label{a-a+}
 \left[ {\hat a^ -  ,\hat a^ +  } \right] = &\frac{{\sqrt 2 }}{{\lambda _0 }}W'\left( x \right).
\end{align}

Analytical expression of superpotential $W\left( x \right)$ can be defined from the action of $\hat a^ -$ to the ground state wavefunction $\psi _0 \left( x \right)$

\[
\hat a^ -  \psi _0 \left( x \right) = 0.
\]

It is the first-order differential equation for $W\left( x \right)$ and its solution leads to

\be
\label{w-x}
W\left( x \right) =  - \frac{1}{{\sqrt 2 \lambda _0 }}\frac{{\frac{d}{{dx}}\psi _0 \left( x \right)}}{{\psi _0 \left( x \right)}}.
\ee

This means that substitution of analytical expression of the ground state wavefunction $\psi _0 \left( x \right)$ of a certain non-relativistic quantum system possessing the canonical approach at  (\ref{w-x}) allows to obtain an analytical expression of the superpotential $W\left( x \right)$.

In the case of the non-relativistic quantum harmonic oscillator model with states bounded at infinity, the potential $V\left( x \right)$ analytically is defined as

\be
\label{v-x}
V\left( x \right) = \frac{{m_0 \omega ^2 x^2 }}{2},\quad  - \infty  < x <  + \infty .
\ee
with $\omega$ being its constant angular frequency. Substitution of (\ref{v-x}) at (\ref{sheq-gen2}) still allows to solve exactly this second order differential equation in terms of the equidistant energy spectrum $E_n$ and wavefunctions of the stationary states $\psi_n \left( x \right)$ of the following analytical form:

\be
\label{en}
E_n  = \hbar \omega \left( {n + \frac{1}{2}} \right),\quad n = 0,1,2, \ldots ,
\ee

\be
\label{wf-on}
\psi _n \left( x \right) = \frac{1}{{\sqrt {2^n n!} }}\left( {\frac{{\lambda_0^2 }}{{\pi }}} \right)^{{\textstyle{1 \over 4}}} e^{ - \frac{1}{2 }{\lambda_0^2 x^2 }} H_n \left( {\lambda_0 x} \right),
\ee
where, $H_n \left( { x} \right)$ are Hermite polynomials being defined in terms of the $_2F_0$ hypergeometric functions~\cite{koekoek2010}.

Oscillator ground state wavefunction of the stationary states $\psi_0 \left( x \right)$ can be easily extracted from Eq.(\ref{wf-on}):

\be
\label{wf-0}
\psi _0 \left( x \right) = \left( {\frac{{\lambda_0^2 }}{{\pi }}} \right)^{{\textstyle{1 \over 4}}} e^{ - \frac{1}{2 }{\lambda_0^2 x^2 }}.
\ee

Then, substitution of (\ref{wf-0}) at (\ref{w-x}) gives us the following analytical expression of $W \left( x \right)$:

\be
\label{we-x}
W\left( x \right) = \frac{1}{{\sqrt 2 }}\lambda _0 x.
\ee

Hence, one obtains exact expressions of the non-relativistic quantum harmonic oscillator creation and annihilation operators $\hat a ^\pm$ through substitution of (\ref{we-x}) at (\ref{a+}) and (\ref{a-})~\cite{dirac1927,messiah1966}:

\begin{align}
\label{a+o}
 \hat a^ +   = \frac{1}{{\sqrt 2 \lambda _0 }}\left( {\lambda _0 ^2 x - \frac{d}{{dx}}} \right), \\
\label{a-o}
 \hat a^ -   = \frac{1}{{\sqrt 2 \lambda _0 }}\left( {\lambda _0 ^2 x + \frac{d}{{dx}}} \right),
\end{align}
as well as an elegant transformation of the commutation relations (\ref{ha+})--(\ref{a-a+}) to the following closed Heisenberg-Weyl algebra of the non-relativistic quantum harmonic oscillator within the canonical approach, which precisely defines the dynamical symmetry of the model under consideration:

\begin{align}
\label{ha+-o}
 \left[ {\hat H,\hat a^ \pm  } \right] = & \pm \hbar \omega \hat a^ \pm  , \\ 
\label{a-a+o}
 \left[ {\hat a^ -  ,\hat a^ +  } \right] = & 1.
\end{align}

Heisenberg-Lie equations can be easily extracted from the above-listed commutation relations as follows:

\begin{align}
\label{hp}
\left[ {\hat H,\hat p_x } \right] = im_0 \hbar \omega ^2 x,\\
\label{hx}
\left[ {\hat H,x} \right] =  - i\frac{\hbar }{{m_0 }}\hat p_x .
\end{align}

The following actions of $\hat a^{\pm}$ operators are valid:

\begin{align}
 \hat a^ -  \psi _n \left( x \right) =& \sqrt n \psi _{n - 1} \left( x \right), \\ 
 \hat a^ +  \psi _n \left( x \right) =& \sqrt {n + 1} \psi _{n + 1} \left( x \right).
\end{align}

Therefore, the oscillator wavefunction of arbitrarily excited $n$ state can easily be recovered as follows:

\be
 \psi _n \left( x \right) = \frac{1}{{\sqrt {n!} }}\left( {\hat a^ +  } \right)^n \psi _0 \left( x \right).
\ee

\section{Construction of a dynamical symmetry algebra of a semiconfined harmonic oscillator model with a position-dependent effective mass}

We are going to follow more or less the same method from the previous section during the initial steps of the construction procedure for a semiconfined oscillator model. However, one needs to take into account that the procedure becomes complicated due to the introduction of the position dependence for the effective mass $m_0 \to M\left(x\right)$ of the quantum system. 

However, one needs to highlight here the fact that the supersymmetric aspects of the position-dependent effective mass concept including the shape-invariance condition have been thoroughly explored in~\cite{bagchi2005}. It uses more general von Roos approach to the kinetic energy operator~\cite{vonroos1983}, therefore the case studied in present paper can be considered by some a way as a particular case of exactly solvable potentials considered in~\cite{bagchi2005}.

Factorization of the position-dependent effective mass Hamiltonian $\hat H$ with the BenDaniel-Duke kinetic energy operator~\cite{bendaniel1966}
\be
\label{h-pdem}
\hat H_0^{BD} =  - \frac{{\hbar ^2 }}{2}\frac{d}{{dx}}\frac{1}{{M\left(x\right)}}\frac{d}{{dx}},
\ee
is well known~\cite{dabrowka1988,cooper1995,plastino1999,gonul2002,dong2007,amir2016}:

\begin{align}
\label{h-a+a-}
\hat H =& \hbar \omega \hat A^ +  \hat A^ -   + E_0 , \\
\label{aa+}
 \hat A^ +   = & - \frac{1}{{\sqrt 2 \lambda _0 }}\frac{d}{{dx}}\sqrt {\frac{{m_0 }}{{M\left( x \right)}}}  + W\left( x \right), \\ 
\label{aa-}
 \hat A^ -   = & \frac{1}{{\sqrt 2 \lambda _0 }}\sqrt {\frac{{m_0 }}{{M\left( x \right)}}} \frac{d}{{dx}} + W\left( x \right).
\end{align}

Here,

\be
\label{h-bd}
\hat H = \hat H_0^{BD} +V \left( x \right).
\ee

Simple computations show that the potential $V\left( x \right)$ can be expressed through the superpotential $W\left( x \right)$ as follows:

\[
V\left( x \right) = E_0  + \hbar \omega W^2 \left( x \right) - \frac{{\hbar \omega }}{{\sqrt 2 \lambda _0 }}\sqrt {\frac{{m_0 }}{{M\left( x \right)}}} \left[ {W'\left( x \right) - \frac{1}{2}W\left( x \right)\frac{{M'\left( x \right)}}{{M\left( x \right)}}} \right].
\]

One can observe that the commutation relations (\ref{ha+})--(\ref{a-a+}) also will have more complicated expressions due to the existence of the mass changing with position:

\begin{align}
\label{haa+}
 \left[ {\hat H,\hat A^ +  } \right] = & \hbar \omega \left\{ {\frac{{\sqrt 2 }}{{\lambda _0 }}\sqrt {\frac{{m_0 }}{{M\left( x \right)}}} W'\left( x \right) + \frac{1}{{4\lambda _0 ^2 }}\frac{{m_0 }}{{M\left( x \right)}}\left[ {\frac{{M''\left( x \right)}}{{M\left( x \right)}} - \frac{3}{2}\left( {\frac{{M'\left( x \right)}}{{M\left( x \right)}}} \right)^2 } \right]} \right\}\hat A^ +   \\ 
  - & \frac{{\hbar \omega }}{{\lambda _0 ^2 }}\frac{{m_0 }}{{M\left( x \right)}}\left[ {W''\left( x \right) - \frac{1}{2}W'\left( x \right)\frac{{M'\left( x \right)}}{{M\left( x \right)}}} \right] \nonumber \\
	-& \frac{{\hbar \omega }}{{4\sqrt 2 \lambda _0 ^3 }}\left( {\frac{{m_0 }}{{M\left( x \right)}}} \right)^{\frac{3}{2}} \left[ {\frac{{M'''\left( x \right)}}{{M\left( x \right)}} - 5\frac{{M''\left( x \right)M'\left( x \right)}}{{M^2 \left( x \right)}} + \frac{9}{2}\left( {\frac{{M'\left( x \right)}}{{M\left( x \right)}}} \right)^3 } \right], \nonumber \\
\label{haa-}
\left[ {\hat H,\hat A^ -  } \right] =&  - \hbar \omega \left\{ {\frac{{\sqrt 2 }}{{\lambda _0 }}\sqrt {\frac{{m_0 }}{{M\left( x \right)}}} W'\left( x \right) + \frac{1}{{4\lambda _0 ^2 }}\frac{{m_0 }}{{M\left( x \right)}}\left[ {\frac{{M''\left( x \right)}}{{M\left( x \right)}} - \frac{3}{2}\left( {\frac{{M'\left( x \right)}}{{M\left( x \right)}}} \right)^2 } \right]} \right\}\hat A^ -  ,\\
\label{aa-aa+}
\left[ {\hat A^ -  ,\hat A^ +  } \right] =& \frac{{\sqrt 2 }}{{\lambda _0 }}\sqrt {\frac{{m_0 }}{{M\left( x \right)}}} W'\left( x \right) + \frac{1}{{4\lambda _0 ^2 }}\frac{{m_0 }}{{M\left( x \right)}}\left[ {\frac{{M''\left( x \right)}}{{M\left( x \right)}} - \frac{3}{2}\left( {\frac{{M'\left( x \right)}}{{M\left( x \right)}}} \right)^2 } \right].
\end{align}

The Hamiltonian (\ref{h-bd}) has a mass changing with position, but it is still positive definite. Therefore, in this case, the following equation

\[
\hat A^ -  \tilde \psi _0 \left( x \right) = 0
\]
also holds. Then, one observes that an exact expression of the superpotential $W\left( x \right)$ can be computed as follows:

\be
\label{w-sc}
W\left( x \right) =  - \frac{1}{{\sqrt 2 \lambda _0 }}\sqrt {\frac{{m_0 }}{{M\left( x \right)}}} \frac{{\frac{d}{{dx}}\tilde \psi _0 \left( x \right)}}{{\tilde \psi _0 \left( x \right)}}.
\ee

The case of a position-dependent effective mass semiconfined harmonic oscillator model with the potential

\be
\label{vm-x}
V\left( x \right) = \frac{{M\left(x\right) \omega^2 x^2 }}{2},\quad  - a  < x <  + \infty ,
\ee
and a position-dependent effective mass

\be
\label{mx}
M\left( x \right) = \begin{cases}
 \frac{{am_0 }}{{x + a}}, & \text{ for } -a<x<+\infty \\
+\infty, &\text{ for } x\leq -a
\end{cases},
\qquad (a>0),
\ee
leads to the exact solution of the stationary Schr\"odinger equation in terms of the wavefunctions of the stationary states

\begin{align}
\label{wf-full}
\tilde \psi _n \left( x \right) = & C_n\cdot \left( {1 + \frac{x}{a}} \right)^{\lambda_0^2 a^2 } e^{ - \lambda_0^2 a\left( {x + a} \right)} L_n^{\left( {2\lambda_0^2 a^2 } \right)} \left( {2\lambda_0^2 a\left( {x + a} \right)} \right), \\
\label{cn}
C_n  = & \left( { - 1} \right)^n \left( {2\lambda_0^2 a^2 } \right)^{\lambda_0^2 a^2  + \frac{1}{2}} \sqrt {\frac{{n!}}{{a\Gamma \left( {n + 2\lambda_0^2 a^2  + 1} \right)}}} ,
\end{align}
and discrete energy spectrum

\be
\label{en-n}
E_n  = \hbar \omega \left( {n + \frac{1}{2}} \right).
\ee

Physics bases of the choice of the position-dependent mass $M\left( x \right)$ in the analytical form~(\ref{mx}) are directly connected to the recent development of the advanced growing technologies of the nanostructures on the substrate as well as with recent few nanometers scale achievements for the MOSFET structures directly exhibiting semiconfined quantum effects~\cite{brune1998,utama2013,bae2018}.

The main feature of the model under consideration is that its energy spectrum completely overlaps with the non-relativistic quantum harmonic oscillator energy spectrum~(\ref{en}), whereas the wavefunction (\ref{wf-full}) is expressed through the generalized Laguerre polynomials $L_n^{\left( \alpha  \right)} \left( x \right)$~\cite{koekoek2010}.

Computation of the superpotential $W\left(x \right)$ leads to its following analytical expression:

\be
\label{spsc}
W\left( x \right) =  \frac{{\lambda _0 a}}{{\sqrt 2 }}\left( {\sqrt {\frac{{m_0 }}{{M\left( x \right)}}}  - \sqrt {\frac{{M\left( x \right)}}{{m_0 }}} } \right).
\ee

Its substitution at (\ref{aa+}) and (\ref{aa-}) gives

\begin{align}
\label{aa+sc}
 \hat A^ +   = & \frac{1}{{\sqrt 2 \lambda _0 }}\left[\lambda _0^2 a\left( {\sqrt {\frac{{m_0 }}{{M\left( x \right)}}}  - \sqrt {\frac{{M\left( x \right)}}{{m_0 }}} } \right)- \frac{d}{{dx}}\sqrt {\frac{{m_0 }}{{M\left( x \right)}}}\right], \\ 
\label{aa-sc}
 \hat A^ -   = & \frac{1}{{\sqrt 2 \lambda _0 }} \left[ \lambda _0^2 a\left( {\sqrt {\frac{{m_0 }}{{M\left( x \right)}}}  - \sqrt {\frac{{M\left( x \right)}}{{m_0 }}} } \right)+\sqrt {\frac{{m_0 }}{{M\left( x \right)}}} \frac{d}{{dx}} \right].
\end{align}

Now, the following commutation relation of these two operators holds:

\be
\label{aa-aa+sc}
\left[ {\hat A^ -  ,\hat A^ +  } \right] = 1 - \frac{1}{{2\sqrt 2 \lambda _0 a}}\sqrt {\frac{{M\left( x \right)}}{{m_0 }}} \left( {\hat A^ +   + \hat A^ -  } \right).
\ee

This expression can be extracted from (\ref{aa-aa+}) by substitution of (\ref{mx}) and (\ref{spsc}) at RHS of it.

Taking into account that

\begin{align}
\label{haa+sc}
\left[ {\hat H,\hat A^ +  } \right] = \hbar \omega \hat A^ +  \left[ {\hat A^ -  ,\hat A^ +  } \right],\\
\label{haa-sc}
\left[ {\hat H,\hat A^ -  } \right] =  - \hbar \omega \left[ {\hat A^ -  ,\hat A^ +  } \right]\hat A^ -  ,
\end{align}
one writes down these commutation relations as follows:

\begin{align}
\label{haa+sc2}
\left[ {\hat H,\hat A^ +  } \right] = & \hbar \omega \hat A^ +   - \frac{{\hbar \omega }}{{2\sqrt 2 \lambda _0 a}}\sqrt {\frac{{M\left( x \right)}}{{m_0 }}} \left[ {\hat A^ +  \hat A^ -   + \left( {\hat A^ +  } \right)^2 } \right] + \frac{{\hbar \omega }}{{8\lambda _0 ^2 a}}\frac{{M'\left( x \right)}}{{M\left( x \right)}}\left( {\hat A^ +   + \hat A^ -  } \right),\\
\label{haa-sc2}
\left[ {\hat H,\hat A^ -  } \right] = & - \hbar \omega \hat A^ -   + \frac{{\hbar \omega }}{{2\sqrt 2 \lambda _0 a}}\sqrt {\frac{{M\left( x \right)}}{{m_0 }}} \left[ {\hat A^ +  \hat A^ -   + \left( {\hat A^ -  } \right)^2 } \right].
\end{align}

One observes that an introduction of operators (\ref{aa+sc}) and (\ref{aa-sc}) does lead yet to close dynamical symmetry algebra that could describe the semiconfined model under consideration. At same time, taking into account that an introduction of the position-dependent mass concept does not violate the commutation relation $\left[ {x,\hat p_x} \right] = i\hbar$, it is interesting to understand how this concept violates the Heisenberg-Lie equations (\ref{hp}) and (\ref{hx}). Straightforward computations lead to the following expressions:

\begin{align}
\label{hpsc}
 \left[ {\hat H,\hat p_x } \right] = & i\hbar M\left( x \right)\omega ^2 x\left( {1 + \frac{1}{{2a}}\frac{{M\left( x \right)}}{{m_0 }}x} \right) + i\hbar \frac{1}{{2am_0 }}\hat p_x ^2 , \\ 
\label{hxsc}
 \left[ {\hat H,x} \right] = & - i\frac{\hbar }{{M\left( x \right)}}\left( {\hat p_x  - \frac{{i\hbar }}{{2a}}\frac{{M\left( x \right)}}{{m_0 }}} \right).
\end{align}

Let's introduce a new operator

\be
\label{gp}
\hat P_x  = \frac{a}{\hbar }\frac{{m_0 }}{{M\left( x \right)}}\hat p_x .
\ee

Then, the commutation relation between the Hamiltonian $\hat H$ and $\hat P_x$ gives

\be
\label{hpsc2}
\left[ {\hat H,\hat P_x } \right] = i\left( {m_0 \omega ^2 ax - \hat H} \right),
\ee
as well as substitution of (\ref{gp}) at (\ref{hxsc}) leads to its following mathematically simplified form:

\be
\label{hxsc2}
\left[ {\hat H,x} \right] =  - \frac{{\hbar ^2 }}{{am_0 }}\left( {i\hat P_x  + \frac{1}{2}} \right).
\ee

Then, as a result of commutation relations (\ref{hpsc2}) and (\ref{hxsc2}) the following three generators $K_0$, $K_1$ and $K_2$ can be introduced:

\begin{align}
\label{k0}
 K_0  = & \hat A^ +  \hat A^ -   + \lambda _0 ^2 a^2  + \frac{1}{2}, \\ 
\label{k1}
 K_1  = & \lambda _0 ^2 ax - \hat A^ +  \hat A^ -   - \frac{1}{2}, \\ 
\label{k2}
 K_2  = & \frac{i}{2} - \hat P_x  . 
\end{align}

The exact analytical expressions of all three generators are as follows:

\begin{align}
 K_0  =& \frac{{\lambda _0 ^2 a}}{2}\left( {x + a + \frac{{a^2 }}{{x + a}}} \right) - \frac{1}{{2\lambda _0 ^2 a}}\frac{d}{{dx}}\left( {x + a} \right)\frac{d}{{dx}}, \\ 
 K_1  = &\frac{{\lambda _0 ^2 a}}{2}\left( {x + a - \frac{{a^2 }}{{x + a}}} \right) + \frac{1}{{2\lambda _0 ^2 a}}\frac{d}{{dx}}\left( {x + a} \right)\frac{d}{{dx}}, \\ 
 K_2  = &i\left[ {\frac{1}{2} + \left( {x + a} \right)\frac{d}{{dx}}} \right].
\end{align}

One can easily check that these generators form closed $\su\left(1,1\right)$ Lie algebra, i.e. the following commutation relations between them hold:

\begin{align}
 \left[ {K_0 ,K_1 } \right] = & iK_2 , \nonumber \\ 
\label{su}
 \left[ {K_2 ,K_0 } \right] = & iK_1 , \\ 
 \left[ {K_1 ,K_2 } \right] = & -iK_0 . \nonumber 
\end{align}

Introducing the following two new operators $K_-$ and $K_+$

\begin{align}
\label{k-}
 K_ -   = K_1  - iK_2 , \\ 
\label{k+}
 K_ +   = K_1  + iK_2 ,
\end{align}
one obtains the following commutation relations between them and $K_0$:

\begin{align}
\label{k-+}
 \left[ {K_ -  ,K_ +  } \right] =& 2K_0 , \\ 
\label{k0k+-}
 \left[ {K_0 ,K_ \pm  } \right] = & \pm K_ \pm  . 
\end{align}

Hermiticity of $K_1$ and $K_2$ as well as $K_\pm$ operators easily can be proven, too. 

The following actions are also valid:

\begin{align}
 K_-\tilde \psi _n \left( x \right)  =& \varepsilon _n \tilde \psi _{n - 1} \left( x \right),  \\ 
 K_ + \tilde \psi _n \left( x \right)  =& \varepsilon _{n + 1} \tilde \psi _{n+1} \left( x \right),
\end{align}
where, 
\be
 \varepsilon _n  =  - \sqrt {n\left( {n + 2\lambda_0^2a^2 } \right)}.
\ee

Thus, one writes down that
\be
 \tilde \psi _n \left( x \right) = \frac{{\left( { - 1} \right)^n }}{{\sqrt {n!\left( {2\lambda_0^2a^2  + 1} \right)_n } }}\left( {K_ +  } \right)^n \tilde \psi _0 \left( x \right). 
\ee

Obtaining the exact expressions of all three generators of the $\su\left(1,1\right)$ Lie algebra, corresponding to the semiconfined oscillator model under consideration, we achieved our main goal. We are going to discuss briefly generalization of the well-known uncertainty principle in the case of the replacement of the constant mass with its position varying analogue as well as the limit relations between all three generators of the $\su\left(1,1\right)$ Lie algebra, corresponding to the semiconfined oscillator model under consideration and the generators of the ordinary quantum harmonic oscillator in a final section.

\section{Conclusions}

Taking into account that the close $\su\left(1,1\right)$ Lie algebra of three generators is constructed and we have their exact analytical expressions, it is necessary to prove now their correctness. One needs to explore their behavior under the limit $a\to \infty$. This is the known case when semiconfinement effect disappears and the quantum system under consideration becomes a non-relativistic quantum harmonic oscillator model within the canonical approach with the wavefunctions of the stationary states expressed through the Hermite polynomials.

First of all, let's discuss briefly the limit that reduces commutation relations (\ref{haa+})--(\ref{aa-aa+}) to their commutation relation analogues (\ref{ha+})--(\ref{a-a+}) in case of the $a \to \infty$. Correctness of limit relations for them can easily be proven due to that the position-dependent mass $M\left( x \right)$ becomes as a constant mass $m_0$ under this limit, hence any derivative of the position-dependent mass $M\left( x \right)$ appeared in expression of commutation relations (\ref{haa+})--(\ref{aa-aa+}) becomes zero. Then, one observes complete recovery of commutation relations (\ref{ha+})--(\ref{a-a+}).

It is appropriate to choose the limit of factorization operators $\hat A ^\pm$ (\ref{aa+sc}) and (\ref{aa-sc}) to their non-relativistic oscillator analogues $\hat a ^\pm$ (\ref{a+o}) and (\ref{a-o}) as a next point of our limit computations. Let's write down exact analytical expressions of $\hat A ^\pm$ taking into account $M\left( x \right)$ from (\ref{mx}):

\begin{align}
\label{aa+sc2}
 \hat A^ +   = \frac{1}{{\sqrt 2 \lambda _0 }}\left[ {\lambda _0 ^2 a\left( {\sqrt {\frac{{x + a}}{a}}  - \sqrt {\frac{a}{{x + a}}} } \right) - \frac{d}{{dx}}\sqrt {\frac{{x + a}}{a}} } \right], \\ 
\label{aa-sc2}
 \hat A^ -   = \frac{1}{{\sqrt 2 \lambda _0 }}\left[ {\lambda _0 ^2 a\left( {\sqrt {\frac{{x + a}}{a}}  - \sqrt {\frac{a}{{x + a}}} } \right) + \sqrt {\frac{{x + a}}{a}} \frac{d}{{dx}}} \right]. 
\end{align}

Here one needs to take into account that

\[
\sqrt {\frac{{x + a}}{a}}  - \sqrt {\frac{a}{{x + a}}}  = \frac{1}{{\sqrt a }}\frac{x}{{\sqrt {x + a} }},
\]
and the following Taylor expansion holds:

\[
\sqrt {\frac{{x + a}}{a}}  = \sqrt {1 + \frac{x}{a}}  = 1 + \frac{x}{{2a}} - \frac{{x^2 }}{{8a^2 }} +  \ldots .
\]

Then, the correctness of the limit of factorization operators $\hat A ^\pm$ (\ref{aa+sc2}) and (\ref{aa-sc2}) to their non-relativistic oscillator analogues $\hat a ^\pm$ (\ref{a+o}) and (\ref{a-o}) will easily be proven.

Taking into account that the generator $K_0$ (\ref{k0}) is a Hamiltonian of the position-dependent mass quantum system under consideration shifted by the parameter $\lambda_0^2 a^2$, the following limit definitely holds:

\[
\mathop {\lim }\limits_{a \to \infty } \left( {K_0  - \lambda _0 ^2 a^2 } \right) =  - \frac{1}{{2\lambda _0 ^2 }}\frac{{d^2 }}{{dx^2 }} + \frac{1}{2}\lambda _0 ^2 x^2 ,
\]
as well as the following limit relations for two other generators $K_1$ (\ref{k1}) and $K_2$ (\ref{k2}) can be derived, too:

\begin{align*}
 \mathop {\lim }\limits_{a \to \infty } \frac{{K_1 }}{{\sqrt 2 \lambda _0 a}} =& \frac{1}{{\sqrt 2 }}\lambda _0 x, \\ 
 \mathop {\lim }\limits_{a \to \infty } \frac{{K_2 }}{{\sqrt 2 \lambda _0 a}} = &\frac{i}{{\sqrt 2 \lambda _0 }}\frac{d}{{dx}}.
\end{align*}

As a consequence of these limit relations, one observes for generators $K_-$ (\ref{k-}) and $K_+$ (\ref{k+}) that

\[
\mathop {\lim }\limits_{a \to \infty } \frac{{K_ \pm  }}{{\sqrt 2 \lambda _0 a}} = \hat a^ \pm  .
\]

Correct recovery of non-relativistic constant-mass harmonic oscillator wavefunctions $\psi _n \left( x \right)$ (\ref{wf-on}) from the position-dependent mass oscillator wavefunctions $\tilde \psi _n \left( x \right)$ (\ref{wf-full}) was already discussed briefly in~\cite{jafarov2021}. One needs to highlight here that it is necessary to use the following known limit between the generalized Laguerre and Hermite polynomials~\cite{koekoek2010}:

\[
\mathop {\lim }\limits_{\alpha  \to +\infty } \left( {\frac{2}{\alpha }} \right)^{\frac{1}{2}n} L_n^{\left( \alpha  \right)} \left( {\left( {2\alpha } \right)^{\frac{1}{2}} x + \alpha } \right) = \frac{{\left( { - 1} \right)^n }}{{n!}}H_n \left( x \right).
\]
as well as the following asymptotical relations:

\begin{align*}
 \left( {2\lambda _0 ^2 a^2 } \right)^{\lambda _0 ^2 a^2  + \frac{1}{2}}  \cong & \sqrt 2 \lambda _0 ae^{\lambda _0 ^2 a^2 \ln 2\lambda _0 ^2 a^2 } , \\ 
 \frac{1}{{\sqrt {\Gamma \left( {n + 2\lambda _0 ^2 a^2  + 1} \right)} }} \cong & \frac{1}{{\sqrt {2\lambda _0 a\sqrt \pi  } }}e^{\lambda _0 ^2 a^2  - \left( {\lambda _0 ^2 a^2  + \frac{n}{2}} \right)\ln 2\lambda _0 ^2 a^2 } , \\ 
 \left( {\frac{{x + a}}{a}} \right)^{\lambda _0 ^2 a^2 }  \cong & e^{\lambda _0 ^2 \left( {ax - \frac{{x^2 }}{2}} \right)} .
\end{align*}

Their substitution at (\ref{wf-full}) correctly recovers (\ref{wf-on}) under the limit $a \to \infty$.

Let's also briefly discuss possible generalized uncertainty principle aspects within the framework of the appearance of the position dependence of the mass of the model. It is well known that the mean values of the non-relativistic quantum harmonic oscillator position and momentum in stationary states can be easily computed through wavefunctions~(\ref{wf-on}) and they are zero:

\be
\label{xn-pn}
\bar x_n  = \int\limits_{ - \infty }^\infty  {x \cdot \psi _n ^2 \left( x \right)dx} = 0 ,\quad \left( {\bar p_x } \right)_n  = \int\limits_{ - \infty }^\infty  {\psi _n \left( x \right)\hat p_x \psi _n \left( x \right)dx} = 0.
\ee

But, one can also find that

\[
\int\limits_{ - \infty }^\infty  {x^2  \cdot \psi _n ^2 \left( x \right)dx}  = \lambda _0 ^{ - 2} \left( {n + \frac{1}{2}} \right),\quad \int\limits_{ - \infty }^\infty  {\psi _n \left( x \right)\hat p_x ^2 \psi _n \left( x \right)dx}  = \hbar ^2 \lambda _0 ^2 \left( {n + \frac{1}{2}} \right).
\]

Therefore, the variances of position and momentum give

\[
\sigma _x ^2  = \int\limits_{ - \infty }^\infty  {x^2  \cdot \psi _n ^2 \left( x \right)dx}  - \left(\bar x_n \right)^2 = \lambda _0 ^{ - 2} \left( {n + \frac{1}{2}} \right),
\]
\[
\sigma _{p_x } ^2  = \int\limits_{ - \infty }^\infty  {\psi _n \left( x \right)\hat p_x ^2 \psi _n \left( x \right)dx}  - \left( {\bar p_x } \right)_n^2  = \hbar ^2 \lambda _0 ^2 \left( {n + \frac{1}{2}} \right).
\]

Their product results in the following uncertainty principle for the non-relativistic quantum harmonic oscillator model with a constant mass:

\[
\sigma _x ^2 \sigma _{p_x } ^2  = \hbar ^2 \left( {n + \frac{1}{2}} \right)^2 .
\]

Let's now adopt these results to the case of the semiconfined oscillator model with the wavefunctions of the arbitrary $n$ stationary states~(\ref{wf-full}). Straightforward computations show that the following mean value of the position different than zero can be obtained:

\[
\bar x_n  = \int\limits_{ - a }^\infty  {x \cdot \tilde \psi _n ^2 \left( x \right)dx}  = \frac{1}{{\lambda _0 ^2 a}}\left( {n + \frac{1}{2}} \right).
\]

Its value being different than zero is due to the asymmetrical behavior of the semiconfinement potential~(\ref{vm-x}) and the mass~(\ref{mx}), but the zero value of $\bar x_n$ from~(\ref{xn-pn}) will be easily recovered under the following limit:

\[
\mathop {\lim }\limits_{a \to \infty } \bar x_n  = 0.
\]

At the same time, the mean value of the momentum does not differ from~(\ref{xn-pn}) and it is simply zero as follows:

\[
\left( {\bar p_x } \right)_n  = \int\limits_{ - a}^\infty  {\tilde \psi _n \left( x \right)\hat p_x \tilde \psi _n \left( x \right)dx}  = 0.
\]

Next, the mean values of the square of the position and momentum give

\[
\int\limits_{ - a}^\infty  {x^2  \cdot \tilde\psi _n ^2 \left( x \right)dx}  = \lambda _0 ^{ - 2} \left( {n + \frac{1}{2}} \right) + \frac{3}{2}\frac{{n\left( {n + 1} \right) + 1}}{{\lambda _0 ^4 a^2 }},
\]
\[
\int\limits_{ - a}^\infty  {\tilde \psi _n \left( x \right)\hat p_x ^2 \tilde \psi _n \left( x \right)dx}  = \frac{1}{2}\hbar ^2 \lambda _0 ^2 \alpha \left[ {\alpha ^2 \frac{{n!}}{{\Gamma \left( {n + \alpha  + 1} \right)}}\sum\limits_{k = 0}^n {\left( {k + 1} \right)^2 \frac{{\Gamma \left( {n + \alpha  - k - 1} \right)}}{{\left( {n - k} \right)!}}}  - 1} \right],
\]
where $\alpha = 2 \lambda _0 ^2a^2$.

One can easily check that the following correct limits of the mean values of the square of the position and momentum hold:

\[
\mathop {\lim }\limits_{a \to \infty } \int\limits_{ - a}^\infty  {x^2  \cdot \tilde \psi _n ^2 \left( x \right)dx}  = \lambda _0 ^{ - 2} \left( {n + \frac{1}{2}} \right),
\]
\[
\mathop {\lim }\limits_{a \to \infty } \int\limits_{ - a}^\infty  {\tilde \psi _n \left( x \right)\hat p_x ^2 \tilde \psi _n \left( x \right)dx}  = \hbar ^2 \lambda _0 ^2 \left( {n + \frac{1}{2}} \right).
\]

Then, the variances of position and momentum now give

\[
\sigma _x ^2  = \int\limits_{ - a}^\infty  {x^2  \cdot \tilde \psi _n ^2 \left( x \right)dx}  - \left( {\bar x_n } \right)^2  = \lambda _0 ^{ - 2} \left( {n + \frac{1}{2}} \right) + \frac{{2n\left( {n + 1} \right) + 1}}{{4\lambda _0 ^4 a^2 }},
\]
\[
\sigma _{p_x } ^2  = \int\limits_{ - \infty }^\infty  {\tilde \psi _n \left( x \right)\hat p_x ^2 \tilde \psi _n \left( x \right)dx}  - \left( {\bar p_x } \right)_n ^2  = \frac{1}{2}\hbar ^2 \lambda _0 ^2 \alpha \left[ {\alpha ^2 \frac{{n!}}{{\Gamma \left( {n + \alpha  + 1} \right)}}\sum\limits_{k = 0}^n {\left( {k + 1} \right)^2 \frac{{\Gamma \left( {n + \alpha  - k - 1} \right)}}{{\left( {n - k} \right)!}}}  - 1} \right].
\]

Hence, the uncertainty principle generalized through the position-dependent mass converting ordinary non-relativistic oscillator to semiconfined one is equal to

\[
\sigma _x ^2 \sigma _{p_x } ^2  = \frac{{\hbar ^2 }}{2}\alpha \left[ {n + \frac{1}{2} + \frac{{2n\left( {n + 1} \right) + 1}}{{4\lambda _0 ^2 a^2 }}} \right]\left[ {\alpha ^2 \frac{{n!}}{{\Gamma \left( {n + \alpha  + 1} \right)}}\sum\limits_{k = 0}^n {\left( {k + 1} \right)^2 \frac{{\Gamma \left( {n + \alpha  - k - 1} \right)}}{{\left( {n - k} \right)!}}}  - 1} \right].
\]

This expression is a little complicated for an arbitrary excited state, however, some interesting assessment procedures can be performed for a simplified $n=0$ ground state, which leads to

\[
\sigma _x ^2 \sigma _{p_x } ^2  = \frac{{\hbar ^2 }}{4}\left( {1 + \frac{2}{{2\lambda _0 ^2 a^2  - 1}}} \right).
\]

From this expression one obtains that $\sigma _x ^2 \sigma _{p_x } ^2 > \frac{{\hbar ^2 }}{4}$ is satisfied only if $a > \sqrt {\frac{\hbar }{{2m_0 \omega }}}$.

We also want to list below before unknown to us seven integral relations for the Laguerre polynomials and one summation formula for the Gamma function, which have been obtained during all computations done in the framework of the generalization of the uncertainty principle:

\[
\int\limits_0^\infty  {x^{\alpha  - 1} e^{ - x} \left[ {L_n^{\left( \alpha  \right)} \left( x \right)} \right]^2 dx}  = \sum\limits_{k = 0}^n {\frac{{\Gamma \left( {k + \alpha } \right)}}{{k!}}} ,
\]
\[
\int\limits_0^\infty  {x^{\alpha  - 2} e^{ - x} \left[ {L_n^{\left( \alpha  \right)} \left( x \right)} \right]^2 dx}  = \sum\limits_{k = 0}^n {\left( {k + 1} \right)^2 \frac{{\Gamma \left( {n + \alpha  - k - 1} \right)}}{{\left( {n - k} \right)!}}} ,
\]
\[
\int\limits_0^\infty  {x^{\alpha  - 1} e^{ - x} L_n^{\left( \alpha  \right)} \left( x \right)\frac{d}{{dx}}L_n^{\left( \alpha  \right)} \left( x \right)dx}  = \frac{1}{2}\sum\limits_{k = 0}^n {\left[ {\frac{{\Gamma \left( {k + \alpha } \right)}}{{k!}} - \left( {\alpha  - 1} \right)\left( {k + 1} \right)^2 \frac{{\Gamma \left( {n + \alpha  - k - 1} \right)}}{{\left( {n - k} \right)!}}} \right]} ,
\]
\[
\int\limits_0^\infty  {x^\alpha  e^{ - x} L_n^{\left( \alpha  \right)} \left( x \right)\frac{d}{{dx}}L_n^{\left( \alpha  \right)} \left( x \right)dx}  = 0,
\]
\[
\int\limits_0^\infty  {x^\alpha  e^{ - x} L_n^{\left( \alpha  \right)} \left( x \right)\frac{{d^2 }}{{dx^2 }}L_n^{\left( \alpha  \right)} \left( x \right)dx}  = 0,
\]
\[
\int\limits_0^\infty  {x^{\alpha  + 1} e^{ - x} \left[ {L_n^{\left( \alpha  \right)} \left( x \right)} \right]^2 dx}  = \left( {2n + \alpha  + 1} \right)\frac{{\Gamma \left( {n + \alpha  + 1} \right)}}{{n!}},
\]
\[
\int\limits_0^\infty  {x^{\alpha  + 2} e^{ - x} \left[ {L_n^{\left( \alpha  \right)} \left( x \right)} \right]^2 dx}  = \left[ {6n\left( {n + \alpha  + 1} \right) + \left( {\alpha  + 1} \right)\left( {\alpha  + 2} \right)} \right]\frac{{\Gamma \left( {n + \alpha  + 1} \right)}}{{n!}},
\]
\[
\sum\limits_{k = 0}^n {\frac{{\Gamma \left( {k + \alpha } \right)}}{{k!}}}  = \frac{{\Gamma \left( {n + \alpha  + 1} \right)}}{{\alpha \cdot n!}}.
\]

Concluding, one needs to say that our main goal is achieved and we succeeded with the construction of the dynamical symmetry algebra for a semiconfined harmonic oscillator model with a position-dependent effective mass. We have found three basis elements of the algebra and observed that the algebra defined through those basis elements is a $\su\left(1,1 \right)$ Heisenberg-Lie algebra. In brief, some special cases and the $a \to \infty$ limit relations from the basis elements to the well-known Heisenberg-Weyl algebra of the non-relativistic quantum harmonic oscillator are also discussed. We did not consider a more general case~\cite{jafarov2022} that discusses the same model but under the action of the applied external homogeneous field. However, one wants to note that the construction method discussed in this paper can be trivially applied to the case of the external homogeneous field existence, too.

\section*{Acknowledgement}

This work was supported by the Azerbaijan Science Foundation –- Grant Nr \textbf{AEF-MCG-2022-1(42)-12/01/1-M-01}.

\end{document}